\documentstyle[12pt, fleqn]{article}

\begin{document}
\title{Existence of the magnetization plateau in a class of
exactly solvable Ising-Heisenberg chains}
\author{Jozef Stre\v{c}ka$^1$  and
Michal Ja\v{s}\v{c}ur$^2$ \\
\normalsize Department of Theoretical Physics and Astrophysics,
Faculty of Science, \\
\normalsize P. J. \v{S}af\'{a}rik University,
Moyzesova 16, 041  54 Ko\v{s}ice, Slovak Republic \\
\normalsize E-mail addresses: jozkos@pobox.sk$^1$, jascur@kosice.upjs.sk$^2$}
\date{Submiteed: \today}
\maketitle
\begin{abstract}
The mapping transformation technique is applied to obtain
exact results for the spin-1/2 and spin-$S$ ($S = 1/2, 1$)
Ising-Heisenberg antiferromagnetic chain in a presence of
external magnetic field. Within this scheme,
a field-induced first-order metamagnetic phase transition resulting in
multiplateau magnetization curves, is investigated
in detail. It is found that the scenario of the plateau
formation depends fundamentally on the ratio
between Ising and Heisenberg interaction parameters, as well as
on the XXZ Heisenberg exchange anisotropy strength.
\end{abstract}
PACS:05.50.+q, 75.10.Jm, 75.10.Hk
\newline
{\it Keywords:} Ising-Heisenberg model; Exact solution;
     Quantum antiferromagnetism

\section{Introduction}

Quantum antiferromagnetism in lower dimensions is one of the most
fascinating subjects in the condensed matter physics. In
particular, the antiferromagnetic quantum Heisenberg chains (AFQHC)
with small spins have attracted much attention on account of the
rich quantum behaviour they display. Nevertheless, due to
the strong quantum fluctuations the classical N\'eel state is
not more an eigenstate of the Hamiltonian and thus, more interesting
quantum phases should be expected to occur in the ground
state. The nature of these phases, however, basically depends on
the spin value of atoms. In fact, as conjectured Haldane \cite{[1]}
in 1983, the integer spin AFQHC have a finite energy gap between the ground
state and the first excited state, while the half-odd integer
ones possess a gapless excitation spectrum.

Another striking feature of the AFQHC is the
appearance of fractional magnetization plateaus in the
magnetization process. Extending the original
Lieb-Schultz-Mattis theorem \cite{[2]} Oshikawa, Yamanaka and Affleck
(OYA) \cite{[3]} argued
that the magnetization per site $m$ can be topologically quantized as:
$p (S_u - m) = \mbox{integer},$
where $p$ is a period of ground state in the thermodynamic
limit and $S_u$ denotes a total spin of an elementary unit.
However, this condition
represents just the necessary condition for the plateau-state
formation and does not directly prove its existence.
It is therefore of interest to investigate how the plateau state
is related to the periodicity of specific model \cite{[4]}.
Moreover, from the theoretical viewpoint the plateau state
can also be regarded as a spin-gap state. Hence, the zero-temperature
magnetization curves with plateaus bring an insight into
the ground-state properties of the system, since
the field-induced spin gaps reflect the gapped excitation spectrum.

Despite of an extensive theoretical effort focused on AFQHC,
there still exist only few exactly solvable models
with pure Heisenberg exchange interactions \cite{[5]}, especially for
mixed-spin chains \cite{deVega}. On the other hand, an exact solution for the chains
with alternating Ising- and Heisenberg-type exchange
interactions can be attained in less sophisticated manner.
Indeed, the exact solution for the Ising-Heisenberg bond alternating
chain (originally proposed and solved
by Lieb {\it et al} \cite{[2]}), has been recently successfully generalized
to the case of anisotropic Heisenberg interaction \cite{[6]}.
In order to avoid mathematical complexities connected with the
noncommutability of relevant spin operators, we will introduce in this
article another class of the Ising-Heisenberg chains (with period $p=3$),
that can be treated exactly within the mapping transformation
method. However, the considered model naturally enables to
analyse in detail the mutual competition between Ising- and Heisenberg-type
interactions and moreover, it also proves to be very useful in view of
the confirmation of multiplateau magnetization curves
by an exact calculation.

This paper is organized as follows. In Sec. 2 the detailed
description of the model, as well as
the fundamental aspects of transformation technique are
presented.
In Sec. 3 we are concerned with the analysis of the most
interesting numerical results for typical spin cases
and finally, in Sec. 4 some concluding remarks are also drawn.

\section{Model and method}

In this article we will study a mixed spin-$1/2$ and spin-$S$
($S = 1/2, 1$) Ising-Heisenberg chain in a presence of
external magnetic field. The structure of the considered mixed-spin chain
is depicted in Fig.1,  where the black circles denote the spin-$1/2$ atoms
and the grey ones represent the spin-$S$ atoms. The total Hamiltonian of the
system is given by:
\begin{eqnarray}
\hat {\cal H} =
J \sum_{i, j}
      \bigl [\Delta (\hat S_i^x \hat S_j^x + \hat S_i^y \hat S_j^y)
                                     + \hat S_i^z \hat S_j^z
      \bigr ]
+ J_1 \sum_{k, l} \hat S_k^z \hat \mu_l^z
- H_A \sum_l \hat \mu_l^z - H_B \sum_k \hat S_k^z,
\label{r1}
\end{eqnarray}
where $\hat \mu_l^z$ and $\hat S_k^{\alpha}$ ($\alpha = x, y, z$)
denote the well-known components of standard spin-$1/2$ and spin-$S$
($S = 1/2, 1$) operators, respectively. The parameter $J$
stands for the Heisenberg interaction between nearest-neighbouring
spin-$S$ atoms (the grey atoms) and $\Delta$ is the anisotropy
parameter that allows to control the anisotropic XXZ
interaction between an easy-axis regime
($\Delta < 1$) and an easy-plane regime ($\Delta > 1$).
Furthermore, the interaction parameter $J_1$ describes the
Ising-type exchange interaction between pairs
of nearest-neighbouring spin-$1/2$ and spin-$S$ atoms and finally,
the terms incorporating $H_A$ and $H_B$ respectively, describe the
coupling of spin-$1/2$ atoms and spin-$S$ atoms to
an external magnetic field. As we can see from Fig.1,
all pairs of nearest-neighbouring Heisenberg atoms are surrounded by
Ising-type atoms only, thus
the model under investigation can also be viewed as an Ising model the
bonds of which are doubly decorated by the Heisenberg atoms.
In view of further manipulations, it is useful to rewrite the total
Hamiltonian $\hat {\cal H}$ as a sum of the bond Hamiltonians, i.e.
$\hat {\cal H} = \sum_{k=1}^N {\hat {\cal H}}_k$, where $N$
denotes the total number of Ising-type atoms and the summation runs
over all bonds of the original (undecorated) chain. The bond Hamiltonian
$\hat {\cal H}_k$ contains all the interaction terms associated
with the $k$th couple of Heisenberg atoms (see Fig.1), and it is
given by
\begin{eqnarray}
\hat {\cal H}_k &=&  J \bigl [
  \Delta (\hat S_{k1}^x \hat S_{k2}^x + \hat S_{k1}^y \hat S_{k2}^y)
         + \hat S_{k1}^z \hat S_{k2}^z \bigr ]
  + J_1 (\hat S_{k1}^z \hat \mu_{k1}^z + \hat S_{k2}^z \hat \mu_{k2}^z)
\nonumber \\
       && - H_B (\hat S_{k1}^z + \hat S_{k2}^z)
         - H_A (\hat \mu_{k1}^z + \hat \mu_{k2}^z)/2.
\label{r2}
\end{eqnarray}
Now, exploiting the usual commutation relation for the bond
Hamiltonians (i.e. $[\hat {\cal H}_k, \hat {\cal H}_j] = 0,$ for  $k \neq
j$), the partition function of the system can be partially
factorized, namely,
\begin{eqnarray}
{\cal Z} = \displaystyle \mbox{Tr}_{ \{ \mu \}} \prod_{k=1}^{N}
\displaystyle \mbox{Tr}_{S_{k1}} \displaystyle \mbox{Tr}_{S_{k2}}
\exp(-\beta  \hat {\cal H}_k).
\label{r3}
\end{eqnarray}
In above, $\beta = (k_B T)^{-1}$, $k_B$ being Boltzmann constant and
$T$ the absolute temperature. $\mbox{Tr}_{ \{ \mu \}}$ means a
trace over the degrees of freedom of the Ising spins and
$\mbox{Tr}_{S_{k1}} \mbox{Tr}_{S_{k2}}$
denotes a trace over the $k$th couple of Heisenberg spins.
At this stage one easily observes that the structure of relation (\ref{r3})
implies a possibility to introduce the decoration-iteration mapping
transformation \cite{[61]}
\begin{eqnarray}
\mbox{Tr}_{S_{k1}} \mbox{Tr}_{S_{k2}} \exp(- \beta \hat {\cal H}_k) =
 A~\exp [\beta R \mu_{k1}^z \mu_{k2}^z
          + \beta H_0 (\mu_{k1}^z + \mu_{k2}^z)/2].
\label{r4}
\end{eqnarray}
As usual, the unknown transformation parameters $A$, $R$ and $H_0$ can be
attained by taking into account remaining degrees of freedom of both
Ising spins ($\mu_{k1}$ and $\mu_{k2}$). In this way one obtains
for the transformation parameters $A$, $R$ and $H_0$ the following
expressions
\begin{eqnarray}
       A~= (V_1 V_2 V_3^2 )^{1/4}, \quad
 \beta R = \ln \Bigl( \frac{V_1 V_2}{V_3^2} \Bigr), \quad
 \beta H_0 =  \beta H_A - \ln \Bigl( \frac{V_1}{V_2} \Bigr),
 \label{r5}
\end{eqnarray}
where the  functions
$V_1$, $V_2$ and $V_3$ depend on the spin value of Heisenberg
atoms, as well as on the parameters of Hamiltonian (\ref{r1}) and
they are summarized for both investigated spin cases in the Appendix.

Here, one should emphasize that the mapping relations
(\ref{r4})-(\ref{r5}) enable to transform the
Ising-Heisenberg mixed-spin chain onto a simple spin-1/2
Ising chain with an effective exchange parameter $R$,
placed in an external magnetic field of the magnitude $H_0$.
Indeed, substituting (\ref{r4}) into (\ref{r3}) gives
the following equality
\begin{eqnarray}
{\cal Z} (\beta, J , J_1, \Delta, H_A, H_B) = A^{N} {\cal Z}_0 (\beta, R, H_0),
\label{r6}
\end{eqnarray}
which relates the partition function of Ising-Heisenberg chain ${\cal Z}$
and that one of the spin-$1/2$ Ising chain ${\cal Z}_0$.
Since the explicit expression for ${\cal Z}_0$ is well known \cite{ising},
we can then straightforwardly calculate all relevant thermodynamic  quantities.
For example, the Gibbs free energy ${\cal G}$ of the mixed-spin Ising-Heisenberg
chain is given by
\begin{eqnarray}
{\cal G} = {\cal G}_0 - N k_B T \ln A,
\label{r7}
\end{eqnarray}
where ${\cal G}_0 = - k_B T \ln {\cal Z}_0$ denotes the Gibbs free energy of the
corresponding spin-$1/2$ Ising chain. Next, by differentiating
the Gibbs free energy ${\cal G}$ with respect to $H_A$ and $H_B$
respectively, one directly obtains solution for the total sublattice magnetization.
Of course, other thermodynamic quantities can also be calculated
on the basis of familiar thermodynamic relations, e. g. the
entropy $S$ and the specific heat $C$ can be calculated from
\begin{eqnarray}
S~= - \Bigl( \frac{\partial G}{\partial T}\Bigr)_H,       \hspace{3cm}
C = - T\Bigl( \frac{\partial^2 G}{\partial T^2} \Bigr)_H.
\label{r10}
\end{eqnarray}

Nevertheless, a similar thermodynamic approach cannot be used for
the calculation of other important quantities such as staggered
magnetization, quadrupolar momentum or some correlations.
Fortunately, Eq. (\ref{r6}) in conjunction with the transformation
formula (\ref{r4}) allows after an elementary algebra the derivation
of following exact spin identities \cite{StJa}
\begin{eqnarray}
\langle f_1 (\hat \mu_i^z, \hat \mu_j^z, ... \hat \mu_k^z,) \rangle
\! \! \! &=& \! \! \!  \langle f_1 (\hat \mu_i^z, \hat \mu_j^z, ... \hat \mu_k^z,) \rangle_0,
\nonumber \\
\langle f_2 (\hat S_{k1}^{\alpha}, \hat S_{k2}^{\gamma},
             \hat \mu_{k1}^z, \hat \mu_{k2}^z) \rangle
\! \! \! &=& \! \! \!  \Bigl \langle  \frac{\mbox{Tr}_{S_{k1}} \mbox{Tr}_{S_{k2}} f_2
 (\hat S_{k1}^{\alpha}, \hat S_{k2}^{\gamma}, \hat \mu_{k1}^z, \hat \mu_{k2}^z)
\exp(- \beta \hat {\cal H}_k)}{\mbox{Tr}_{S_{k1}} \mbox{Tr}_{S_{k2}}
\exp(- \beta \hat {\cal H}_k)} \Bigr \rangle,
\label{r8}
\end{eqnarray}
with arbitrary function $f_1$ depending exclusively on Ising spin variables
and the function $f_2$ depending on the spin variables from
the $k$th bond only. The superscript $\alpha, \gamma
\equiv (x, y, z)$ label the spatial components of spin operators
and finally, the symbols $\langle ... \rangle$ and $\langle ... \rangle_0$
stand for the standard ensemble average in the Ising-Heisenberg
and its equivalent simple Ising model, respectively.
However, the above spin identities considerably simplify
the calculation of a large number of quantities.
Indeed, for the reduced sublattice magnetization ($m_A^z$,
$m_B^z$), the total single-site magnetization $m$
and the staggered sublattice magnetization ($m_A^s$, $m_B^s$),
one attains after straightforward algebra
\begin{eqnarray}
m_A^z \! \! \! &\equiv& \! \! \!
\frac12 \langle \hat \mu_{k1}^z + \hat \mu_{k2}^z \rangle
= \frac12 \langle \hat \mu_{k1}^z + \hat \mu_{k2}^z \rangle_0
\equiv  m_0, \nonumber \\
m_B^z \! \! \! &\equiv& \! \! \!
\frac12 \langle \hat S_{k1}^z + \hat S_{k2}^z \rangle =
(V_4/V_1 - V_5/V_2 + 2 V_6/V_3)/2 \nonumber \\
&& - 2 m_0 (V_4/V_1 + V_5/V_2) +
2 \varepsilon_0 (V_4/V_1 - V_5/V_2 - 2 V_6/V_3), \nonumber \\
m \! \! \! &\equiv& \! \! \! (m_A^z + 2 m_B^z)/3, \nonumber \\
m_A^s \! \! \! &\equiv& \! \! \!
\frac12 \langle \hat \mu_{k1}^z - \hat \mu_{k2}^z \rangle
= \frac12 \langle \hat \mu_{k1}^z - \hat \mu_{k2}^z \rangle_0
\equiv  m_0^s, \nonumber \\
m_B^s \! \! \! &\equiv& \! \! \!
\frac12 \langle \hat S_{k1}^z - \hat S_{k2}^z \rangle = - m_0^s V_7/V_3.
\label{r9}
\end{eqnarray}
In above, $m_0$, $m_0^s$ and $\varepsilon_0$ represent the reduced
magnetization, staggered magnetization and nearest-neighbour
correlation of the corresponding undecorated Ising
chain and the coefficients $V_1$-$V_7$ are listed for both
investigated spin cases in the Appendix.

Finally, let us define some pair correlation functions and
the quadrupolar momentum, which are also very useful for
understanding of magnetic properties of the system, namely,
\begin{eqnarray}
q_{hh}^{xx} \! \! \! & \equiv  &  \! \! \! \langle \hat S_{k1}^x \hat S_{k2}^x \rangle \equiv
         \langle \hat S_{k1}^y \hat S_{k2}^y \rangle,
\hspace{1cm}
q_{hh}^{zz}  \equiv  \langle \hat S_{k1}^z \hat S_{k2}^z \rangle,
\hspace{1cm}
q_{ii}^{zz}  \equiv  \langle \hat \mu_{k1}^z \hat \mu_{k2}^z \rangle
             \equiv \varepsilon_0,
\nonumber \\
q_{ih}^{zz} \! \! \! & \equiv & \! \! \!
\frac12 \langle \hat S_{k1}^z \hat \mu_{k1}^z + \hat S_{k2}^z \hat \mu_{k2}^z \rangle ,
\hspace{1.4cm}
\eta \equiv \frac12 \langle (\hat S_{k1}^z)^2 + (\hat S_{k2}^z)^2 \rangle.
\label{r11}
\end{eqnarray}
In these equations, the subscripts denote
the type of atoms and superscripts the space direction.
One should also notice that the definition of the parameter $\eta$
is obviously meaningful for $S \ge 1$ only. Although,
the derivation of relevant equations for these quantities is straightforward,
the calculation procedure by itself is rather lengthy and tedious,
thus the details are not presented here.

\section{Numerical results and discussion}

Before discussing the most interesting numerical
results, it is worth mentioning that some preliminary results for
the ferromagnetic version of the model ($J<0$, $J_1<0$) have
already been published by the present authors elsewhere
\cite{[7]}. For this reason, in this article we will restrict our
attention to the doubly decorated Ising-Heisenberg chain with
antiferromagnetic interactions only (i.e. $J>0, J_1>0$).
The particular attention is focused on the ground-state analysis
and the appearance of plateaus in the chains with different
decorating spin $S$.
Among other matters, we will directly prove the existence of
double-plateau magnetization curve
in the spin $S=1/2$ chain, more precisely, the magnetization
curve with plateaus at $m=0$ and $1/6$.
On the other hand, in the spin $S=1$ chain a greater
diversity of magnetization process will be
confirmed, in fact, we will prove the existence of double-plateau
($m=0$ and $1/2$), triple-plateau ($m=0$, $1/6$ and $1/2$)
and quadruple-plateau ($m=0$, $1/6$, $1/3$ and $1/2$)
magnetization curves.

In addition, since each couple of the Heisenberg
atoms is surrounded by the Ising atoms only, the relevant spin
deviations cannot propagate
through the Ising bonds and therefore, the quantum
fluctuations are necessarily localized within the unit cell
(within the four-spin cluster consisting of the Heisenberg spin pair and its
nearest-neighbouring Ising spins). Owing to
this fact, the observed plateaus cannot be considered as the
OYA plateaus, i. e. the plateaus with collective eigenstate
extending over the whole chain. Nevertheless, it is interesting
to note that all observed fractional magnetization
satisfy the OYA condition for $p=6$
(the period of translational symmetry should be twice as large as
the periodicity of Hamiltonian as a consequence of an
antiferromagnetic nature of ground state).

\subsection{Spin $S=1/2$ chain}

We begin our analysis by considering the effect of exchange anisotropy
$\Delta$ and uniform magnetic field (i.e. $H_A=H_B=H$)
on the ground-state phase boundaries of the spin $S = 1/2$ chain.
For this purpose, we have displayed in Fig.2 some typical ground-state phase diagrams
in the $\Delta-H/J$ plane for $J_1/J = 1.0$ and $2.0$. As one can see from this figure, the relevant phase
boundaries separate three or four distinct phases, namely,
the antiferromagnetic (AF),
ferrimagnetic I~(FI), ferrimagnetic II (FII) and
saturated paramagnetic (SP) phase. One also observes that both
ferrimagnetic phases (FI and FII) represent an intermediate
phase between AF and SP phase occurring due to the first-order
metamagnetic transition.
As we have already mentioned before, different phases can be
distinguished  by analysing the magnetization and the correlation
functions at $T=0$. In this way one finds the following
ground-state results for particular phases: \\
Antiferromagnetic phase (AF):
\begin{eqnarray}
(q_{hh}^{xx}, q_{hh}^{zz}, q_{ih}^{zz}, q_{ii}^{zz}, m_A^s, m_B^s) =
(-J \Delta Q^{-1}, -1/4, -J_1 Q^{-1}, -1/4, 1/2, -2J_1 Q^{-1});
\nonumber
\label{r13}
\end{eqnarray}
where we have defined the function $Q = 4 \sqrt{J_1^2 + (J \Delta)^2}$,
in order to write the relevant expressions in more abbreviated and elegant form. \\
Ferrimagnetic phase I~(FI):
\begin{eqnarray}
(q_{hh}^{xx}, q_{hh}^{zz}, q_{ih}^{zz}, q_{ii}^{zz}, m_A^z, m_B^z, m)
= (-1/4, -1/4, 0, 1/4, 1/2, 0, 1/6);
\nonumber
\label{r14}
\end{eqnarray}
Ferrimagnetic phase II (FII):
\begin{eqnarray}
(q_{hh}^{xx}, q_{hh}^{zz}, q_{ih}^{zz}, q_{ii}^{zz}, m_A^z, m_B^z, m)
= (0, 1/4, -1/4, 1/4, -1/2, 1/2, 1/6);
\nonumber
\label{r15}
\end{eqnarray}
Saturated paramagnetic phase (SP):
\begin{eqnarray}
(q_{hh}^{xx}, q_{hh}^{zz}, q_{ih}^{zz}, q_{ii}^{zz}, m_A^z, m_B^z, m)
= (0, 1/4, 1/4, 1/4, 1/2, 1/2, 1/2).
\nonumber
\label{r16}
\end{eqnarray}
Moreover, it is also noteworthy that all ground-state phase boundaries
can be expressed analytically as follows (see Fig.2), \newline
a) for $J_1/J \leq 1.0$
\begin{eqnarray}
(1)\; H/J = \sqrt{\Delta^2 + (J_1/J)^2} - \Delta, \hspace{1cm}
(2)\; H/J = (\Delta + J_1/J)/2 + 1/2,
\label{11}
\end{eqnarray}
b) $J_1/J > 1.0$
\begin{eqnarray}
\label{12}
(1)\; H/J \! \! \! &=& \! \! \! \sqrt{\Delta^2 + (J_1/J)^2} - \Delta,
\hspace{0.75cm}
(2) \;H/J = (\Delta + J_1/J)/2 + 1/2,       \\
(3) \;H/J \! \! \! &=& \! \! \! \sqrt{\Delta^2 + (J_1/J)^2} - J_1/J + 1, \hspace{0.25cm}
(4) \;H/J = J_1/J, \hspace{0.25cm}
(5) \Delta = J_1/J - 1. \nonumber
\end{eqnarray}
In order to demonstrate the diversity of the magnetization process,
we have plotted in Fig.3 the low-temperature magnetization curves
for various exchange anisotropies $\Delta$. The detailed
examination of these dependencies reveals that the
mechanism of the magnetization process depends basically on the
ratio between Ising and Heisenberg interaction constants,
more specifically, on whether $J_1/J \leq 1.0$ or $J_1/J > 1.0$.
In the former case, the plateau state arises due to the alignment
of the Ising spins towards the external-field direction (FI phase)
regardless of the anisotropy strength $\Delta$ (see Fig.3a).
Naturally, in this case there is no other possibility for the
formation of intermediate plateau.
On the other hand, in the case of $J_1/J>1.0 $ the metamagnetic transition
to the FI phase can be observed for the stronger anisotropies
$\Delta$ only (see Fig.3d), while for the weaker anisotropies
the FII phase becomes the stable one.
In this phase, the Heisenberg (Ising) spins
align parallel (antiparallel) with respect to the external-field
direction as it is apparent from Fig.3b. Moreover, the situation for
the most interesting point at which both intermediate phases
(FI and FII) coexist, is depicted in Fig.3c.
Referring to this plot, the coexistence of both
ferrimagnetic phases is also reflected in the mixed feature
of the magnetization curve (compare magnetization curve from Fig.3c
with those in Figs.3b and 3d).

In order to enable an independent check of the magnetization scenario,
we have also studied the relevant low-temperature dependencies of the
nearest-neighbour correlation functions introduced in Eq. (\ref{r11}).
In Fig.4a, the variations of the correlation functions with
anisotropy $\Delta$ are shown for the system without external
magnetic field. As one can see, the correlation function
$q_{hh}^{zz}$ between Heisenberg spins takes its saturation value
irrespectively of $\Delta$, what means that all
nearest-neighbouring Heisenberg spin pairs align antiparallel
with respect to each other. Moreover, the saturated value of
correlation $q_{ii}^{zz} = -1/4$ indicates a perfect
antiferromagnetic alignment also in the Ising sublattice
(between third nearest-neighbour Ising spins). Contrary to this behaviour,
the perfect antiparallel alignment between Ising and Heisenberg
spins is destroyed as $\Delta$ increases from zero (see the correlation
$q_{ih}^{zz}$). Hence, the Ising and Heisenberg spins are
oriented randomly with respect to each other, the degree of randomness
being the greater, the stronger the exchange anisotropy $\Delta$.
In addition, it is clear that the anisotropy term $\Delta$ is also responsible
for the onset of an interesting short-range ordering (nonzero
$q_{hh}^{xx}$) in $xy$ plane. Anyway, the value of correlation
$q_{hh}^{xx}$ can be thought as a measure of the strength of
local quantum fluctuations appearing in the spin system.
Furthermore, in Figs.4b-4d we have displayed the field dependencies of the
correlation functions corresponding to the magnetization curves
from Figs.3b-3d. The depicted behaviour for the correlation
functions is in complete agreement with results for
the magnetization curves and moreover, it enables better understanding
of the magnetic ordering in relevant phases.
As a typical example we can mention the ferrimagnetic phases.
Although the total magnetization of both the FI and FII phases
is the same, the results for other quantities indicate
a fundamental difference between them. In fact, in the FI phase
the pairs of Heisenberg spins create singlet dimers and thus,
they do not contribute to the total magnetization which is
nonzero due to the fully polarized Ising spins only, as already stated
before. This observation would suggest that the formation
of FI plateau state should be based on the quantum
mechanism with a significant influence of local quantum fluctuations.
In contrast to this, the FII phase represents the standard
ferrimagnetically ordered phase usually observed in the pure Ising systems.
Indeed, the generation of the FII plateau state is nothing
but the gapped excitation from the N\'eel state, implying the
magnetization process with a 'classical' Ising-like mechanism.
Finally, one should notice that after exceeding the saturation field
given by conditions (\ref{11}) and (\ref{12}),
the ground state becomes fully polarized (SP phase) and
all spins are aligned in the external-field direction.

Now, let us investigate the finite-temperature behaviour of the system.
Firstly, we take a closer look at the thermal dependencies of total
magnetization that are shown in Fig.5. As one can  see,
the initial value of total magnetization takes one of three possible
values $m=0$, $1/6$ or $1/2$ for the AF, FI (FII) or SP phase,
respectively. Furthermore, there are two special cases
which correspond to the coexistence of the relevant phases. In these
cases, we have obviously obtained  $m=1/12$ or $1/3$ at $T=0$.
It is also easy to observe here that the most interesting dependencies
appear for external fields from the neighbourhood of
phase boundaries. Hence, the observed rapid increase (decrease)
of the magnetization can be attributed to the thermal excitations of
huge number of spins which occur due to the competitive influence
of both phases.

For completeness, we have also plotted the entropy (Fig.6a) and the
specific heat (Fig.6b-6d) as a function of temperature. As
one can expect, the entropy of the system does not
vanish for the boundary external-field values $H/J=1.0$ and $2.25$,
at which the relevant phases coexist in
the ground state (see Fig.6a). However, we should mention that for any
other external fields the entropy vanishes as the temperature goes to zero.
Another quantity which is interesting also
from the experimental point of view represents the specific heat. The
thermal variations  of this quantity  are depicted in Fig.6b for the same
values of $H/J$ as for entropy in Fig.6a. The displayed behaviour indicates a
round Schottky-type
maximum, whereas the stronger the external field, the flatter and
the broader the maximum. Apart from this trivial finding, one can
also observe the double-peak specific heat curves (see Figs.6c
and 6d).
The first peak which occurs in the specific heat curve at lower
temperature is
evidently closely related to the rapid variation of the
magnetization (compare Fig.6c-d with Fig.5b). Moreover,
it turns out that the maximum of this peak can be located
approximately in the middle of ferrimagnetic region
(in our case around $H/J \approx 1.5$).
On the other hand, the second peak may be thought as a
Schottky-like peak resulting from the antiferromagnetic
short-range order. Indeed, the relevant thermal
dependencies for the nearest-neighbouring correlations
strongly support this statement.

In the following subsection we examine the spin $S=1$ chain
in order to clarify the influence of decorating spin on the
magnetic properties of the system.

\subsection{Spin $S=1$ chain}

We start our discussion once again with the analysis of ground state.
In order to establish correct ground-state phase boundaries
all relevant quantities have been
examined in detail. From this analysis one can conclude that
depending on the ratio between $J_1$, $J (\Delta)$ and $H$, altogether
six different phases can appear in the ground state (see ground-state
phase diagrams in Figs.7a and 9a-b). In this part, we will
firstly describe details of the spin ordering emerging in the appropriate
ground-state phases and then, we will show how the scenario of the
magnetization process depends on the parameters of the model.

In the antiferromagnetic phase (AF) one finds a perfect antiferromagnetic
alignment in the Ising sublattice (i.e. $m_A^s=1/2$ and $q_{ii}^{zz}=-1/4$)
regardless of the strength of the exchange anisotropy $\Delta$.
Accordingly, the relevant spin order in the
Ising sublattice is completely identical to that one in the AF
phase of the spin $S=1/2$ chain. Nevertheless, in contrast to the spin $S=1/2$
case, the antiparallel alignment between nearest-neighbouring Heisenberg
spins is weakened along $z$-axis as $\Delta$ increases from zero.
In fact, the correlation $q_{hh}^{zz}$ tends monotonically from its classical
Ising value $q_{hh}^{zz} = -1$ at $\Delta = 0$ to
$q_{hh}^{zz} = -1/2$ for large $\Delta$ (see e. g. Fig.8a).
In addition, one easily proves also the validity of relation
$\eta = |q_{hh}^{zz}|$ in the whole AF region.
This observation
would suggest that $\Delta$ supports the spin reorientation of
Heisenberg spin pairs, namely, from the antiparallel oriented Heisenberg
spin pair (one spin in $S^z=-1$ state, another one in spin $S^z=1$ state
to be further denoted as the $'1-1'$ spin pair) towards the spin pair
with both Heisenberg spins in the spin $S^z=0$ state (the $'00'$ spin pair).
However, in the large $\Delta$ limit both types of the Heisenberg spin
pairs ($'1-1'$ as well $'00'$) are equally well populated and with a
high probability also randomly distributed among Ising spins.
This suggestion is
strongly supported by results for the correlation $q_{ih}^{zz}$ and
staggered magnetization $m_B^s$ which asymptotically tend to zero
as $\Delta$ $\to$ $\infty$ (Fig.8a). Finally, one should also notice
that all these effects could originate from the onset of the
antiferromagnetic short-range-ordering in the $xy$ plane (nonzero
$q_{hh}^{xx}$) that is the stronger (up to the value $1/ \sqrt{2}$),
the greater the exchange anisotropy strength $\Delta$.

Now, let us turn our attention to both the FI and FII ferrimagnetic
phases. As in the spin $S=1/2$ case, the nonzero
total magnetization of the FI phase arises due to the fully polarized
Ising spins only ($m_A^z = 1/2$ and $q_{ii}^{zz}=1/4$), while the
Heisenberg sublattice does not contribute to the total magnetization
at all ($m_B^z=0$ and $q_{ih}^{zz}=0$).
Since the FI phase can arise by increasing the external field from
the AF phase only (see Fig.7a, 9ab), it is of
interest to compare the relevant spin ordering in both phases. Actually,
one still finds $\eta = |q_{hh}^{zz}|$ to be valid, however,
the nearest-neighbour correlation $q_{hh}^{zz} (q_{hh}^{xx})$
is weaker (stronger) in the FI phase
with respect to that one in the AF phase (Fig.8d). These results are
taken to mean that the number of nearest-neighbouring '00' spin
pairs increases (of course, only up to one half of the total number
of pairs) as one passes through the AF-FI phase boundary.
Moreover, the appearance of a massive short-range order in the $xy$
plane strongly implies the relevance of local quantum fluctuations
also in the FI phase. On the other hand, in the second ferrimagnetic
phase FII we have found the following results
\begin{eqnarray}
(q_{hh}^{xx}, q_{hh}^{zz}, q_{ih}^{zz}, q_{ii}^{zz}, \eta, m_A^z, m_B^z, m)
 = (0, 1, -1/2, 1/4, 1, -1/2, 1, 1/2),
\nonumber
\label{r20}
\end{eqnarray}
indicating the 'classical' character of this phase that
is usually observed also in the pure Ising spin systems.
As the relevant spin-ordering is thoroughly analogous
to that one in the FII phase of spin $S=1/2$ chain, for brevity
we will not repeat its description here.

The most interesting spin order, however, can be found in the
valence-bond phase (VB) and the intermediate phase (IP). Really,
for instance in the VB phase one finds
\begin{eqnarray}
(q_{hh}^{xx}, q_{hh}^{zz}, q_{ih}^{zz}, q_{ii}^{zz}, \eta, m_A^z, m_B^z, m)
 = (- 1/2, 0, 1/4, 1/4, 1/2, 1/2, 1/2, 1/2).
\nonumber
\label{r19}
\end{eqnarray}
As one can see, in contrast to the fully polarized Ising sublattice
($m_A^z=1/2$ and $q_{ii}^{zz}=1/4$),
each couple of the nearest-neighbouring Heisenberg spins consists of one
polarized spin ($S^z=1$) and one spin in the $S^z=0$ state, i. e.
the spin $'01'$ pair.
It is interesting to note that the symmetrization of both Heisenberg spin states
can be achieved using the valence-bond-solid (VBS) picture \cite{[8]}.
Accordingly, each spin-$1$
atom splits into one polarized spin-$1/2$ variable with the fixed projection
into the external field direction and one spin-$1/2$ variable
with the unfixed projection creating a valence-bond.
Thus, it is reasonable to assume that both Heisenberg spins
interchange their spin states (tunneling between the $S^z=0$ and
$S^z=1$ spin states) and therefore, they effectively act on the surrounding
Ising spins as the spin-$1/2$ atoms (see also the result for
correlation $q_{ih}^{zz}$).
However, the antiferromagnetic short-range order in the $xy$ plane
(nonzero $q_{hh}^{xx}$), as well as the fact that the VB phase cannot be detected
in the pure Ising system (i.e. for $\Delta=0$), imply again the
obvious influence of the local quantum fluctuations.

Perhaps the most striking spin alignment has been discovered in
the IP phase. Among other matters, the perfect antiferromagnetic
alignment between the Ising spins ($m_A^s=1/2$ and $q_{ii}^{zz} = -1/4$)
has been confirmed in IP phase, hence, the total magnetization
of system ($m=1/3$) is entirely determined by the contribution
of Heisenberg spins. Anyway, the results for the Heisenberg sublattice magnetization
$m_B^z=1/2$, correlation function $q_{hh}^{zz} = 0$ and quadrupolar momentum
$\eta = 1/2$, clearly indicate that each couple of the
nearest-neighbouring Heisenberg atoms
comprises of the spin $'01'$ pair. However, since both Heisenberg spins
are placed in the IP phase between two non-equivalent Ising spins (one 'up', other
'down'), the spin state interchange between both Heisenberg atoms
would lead to $q_{ih}^{zz}=0$ what is in contradiction with our
numerical result $q_{ih}^{zz} \neq 0$. On the other hand, if the Heisenberg atom
in spin $S^z=1$ state would be strictly antiferromagnetically coupled to
its nearest-neighbouring Ising spin (spin 'down'), then we would have
$q_{ih}^{zz}=-1/4$. Nevertheless, our result for the correlation
$|q_{ih}^{zz}| \ll 1/4$ indicates only the partial antiferromagnetic order
between Ising and Heisenberg spins. This observation would suggest
that the spin $'01'$ Heisenberg pairs should be, to a certain
degree, randomly distributed  among Ising spins. Finally, as one
can expect, in the high field limit the system undergoes a phase
transition towards the fully saturated paramagnetic phase (SP). As before, in
the SP phase all Ising and Heisenberg spins are completely aligned towards
the external-field direction, thus, in the SP phase one attains
\begin{eqnarray}
(q_{hh}^{xx}, q_{hh}^{zz}, q_{ih}^{zz}, q_{ii}^{zz}, \eta, m_A^z, m_B^z, m)
 = (0, 1, 1/2, 1/4, 1, 1/2, 1, 5/6);
\nonumber
\label{r21}
\end{eqnarray}

Now, let us proceed to examine the magnetization process of the
system under investigation. For this purpose,
we have shown in Figs.7a and 9a-b the ground-state phase diagrams
in the $\Delta-H/J$ plane together with some typical examples of the
magnetization curves (Figs.7b-d and 9c-d) for three selected
values of interaction parameters $J_1/J$. In order to provide an independent check of the magnetization
scenario from Figs.7b-d, the corresponding field-dependencies
of the correlation functions and quadrupolar momentum are displayed in
Figs.8b-d. Obviously, all the above results are absolutely in accordance
with the aforementioned ground-state spin ordering.
Moreover, through the comparison of Figs.7a and 9a-b one can
realize that the nature of magnetization process depends basically
on the ratio between the Ising and Heisenberg interaction parameters.
In fact, the stronger the Ising interaction $J_1$ with respect to
the Heisenberg one $J (\Delta)$, the broader the parameter region corresponding
to the 'classical' FII phase. Otherwise, the increasing influence of the Heisenberg
interaction $J (\Delta)$ causes the broadening of regions corresponding
to the FI and VB phase, respectively, until the FII phase completely vanishes
below $J_1/J < 2/3$, see e. g. Fig.9a where the FII phase is already missing.
Consequently, in the FI and
VB phases one can expect that the effect of local quantum fluctuations
plays an important role.

Although, the system exhibits a stepwise magnetization curve with an
abrupt change of the magnetization in the whole range
of parameters (see Figs.7b-d and 9c-d), there is a fundamental difference between the magnetization
process in the 'classical' Ising-like regime and that one in the quantual regime.
As a matter of fact, in the former case one observes the double-plateau
magnetization curve with the FII plateau state only,
i. e. with the FII phase as an intermediate state between the AF and SP phase
(see Fig.7b). Contrary to this,
in the latter case one can detect the double- (Fig.7c),
triple- (Fig.7d), or quadruple-plateau (Fig.9c-d)
magnetization curves. The double-plateau magnetization curves are
rather rarely observable, since they arise due to the direct metamagnetic
transition from the AF phase to the VB phase in a relatively narrow region
of $\Delta$ only (see Fig.7a). However,
when the mechanism of the magnetization process is driven by the Heisenberg
interaction, i. e. under the requirement of sufficiently small
$J_1/J$ and sufficiently large $\Delta$, the triple-plateau magnetization
curves with the transitions between AF-FI-VB-SP phases are always preferred (Fig.7d).
Finally, one should also remark that if the condition $\Delta \approx H/J
\approx J_1/J > 1$ is satisfied, there also appear the
extraordinary quadruple-plateau magnetization curves (see Fig.9b)
with the IP phase in a very narrow region of the external fields.
Indeed, we found even two different possibilities for the quadruple-cascade
transitions, namely, the AF-FI-IP-FII-SP cascade transitions
(Fig.9c) as well as the AF-FI-IP-VB-SP ones (Fig.9d).

To conclude the analysis of the spin $S=1$ chain,
we will also briefly mention the finite-temperature
behaviour of the system under consideration. For this purpose, the thermal
variations of the total magnetization are plotted in Fig.10
for $J_1/J=1.0$ and two selected values of the anisotropy $\Delta$.
In agreement with the aforementioned arguments, one observes here
three and two field-induced transitions for the anisotropy
strengths $\Delta=1.0$ and $0.55$, respectively. Evidently, as
the magnitude of external field varies, the various thermal
dependencies result from the competition between the Ising interaction,
Heisenberg interaction and magnetic field. However,
the most interesting dependencies arise again for the
external fields from the vicinity of the phase boundaries. In
such a case, the relevant thermal excitations result in a very rapid change of
magnetization due to the competing influence of the phases
separated by relevant transition line. Moreover, it turns out that
the narrower interval of external fields corresponds to
the relevant phase, the more robust change in the
magnetization can be observed.

\section{Conclusion}

In the present article we have obtained the exact solution of
the mixed spin-$1/2$ and spin-$S$ ($S=1/2, 1$) Ising-Heisenberg
chain in the external magnetic field.
The most important result stemming from this study is
the confirmation of multistep magnetization process by an exact calculation.
Moreover, it has been proved that the character of magnetization
process depends essentially on the ratio between the Ising and
Heisenberg interactions, whereas the XXZ anisotropy term $\Delta$
also allows to control it in deciding manner. Since the presence of
the non-diagonal interaction term $J \Delta$ is responsible for the
onset of the local quantum fluctuations, as we have also shown, it basically
modifies an otherwise trivial Ising-like behaviour. For example,
in the case of the spin $S=1$ chain one finds instead of the double-plateau
magnetization curve arising in the pure Ising spin system ($\Delta = 0$),
the double-, triple-, or even quadruple-plateau magnetization curves
in the Ising-Heisenberg chain with $\Delta \neq 0$.
Altogether, the presented results indicate that the extraordinary
rich ground-state phase diagrams result from the competition between
the easy-axis interactions $J_1, J$ and the easy-plane interaction $J \Delta$.

One should also emphasize that our research on Ising-Heisenberg chains
has been stimulated by the recent experimental works dealing with
many quasi-1D mixed-spin chains \cite{[9]}. Although, we are not aware
of any quasi-1D system in that two kinds of magnetic ions regularly
alternate with $p=3$ periodic fashion (AABAAB...),
the recent progress in molecular engineering \cite{[10]}
supports our hope that the synthesis of such a polymeric chain
should be possible in the near future. Structural derivatives of
a novel polymeric chain recently reported by Mukherjeee {\it et al}
\cite{muk}, seem to be the most promising candidates from this point of view.
In fact, the crystal structure of the above mentioned polymeric
chain consists of the spin-1/2 $\mbox{Cu}^{{\scriptsize \mbox{II}}}$
dimers linked through $\mbox{Ni}^{{\scriptsize \mbox{II}}}$ monomers.
Unfortunately, as a consequence of the square-planar
coordination of $\mbox{Ni}^{{\scriptsize \mbox{II}}}$ ions in
$[\mbox{Ni}^{{\scriptsize \mbox{II}}} (\mbox{CN})_4]^{2-}$
bridging groups, the $\mbox{Ni}^{{\scriptsize \mbox{II}}}$ metal
ions are diamagnetic and thus, they do not contribute to the magnetism.

Finally, it should be stressed that the applied mapping transformation
technique does not require any restriction to the dimensionality
of the spin system and hence, it can be straightforwardly generalized
to the mixed Ising-Heisenberg lattices in two- and three-dimensions,
too \cite{StJa}.
\newline
{\it Acknowledgement:}
The authors would like to thank referee for useful comments. \\
This work was supported by the VEGA grant No. 1/9034/02
and APVT grant No. 20-009902.
\newline

\section{Appendix}
Explicit expressions for the functions $V_1$-$V_6$.
\newline
a) Spin-1/2 chain:
\begin{eqnarray}
V_1 \! \! \! &=& \! \! \! 2 \exp(- \beta J/4) \cosh(\beta J_1/2 + \beta H_B)
+ 2 \exp(\beta J/4) \cosh(\beta J \Delta/2 ) \nonumber \\
V_2 \! \! \! &=& \! \! \! 2 \exp(- \beta J/4) \cosh(\beta J_1/2 - \beta H_B)
+ 2 \exp(\beta J/4) \cosh(\beta J \Delta/2 ) \nonumber \\
V_3 \! \! \! &=& \! \! \! 2 \exp(- \beta J/4) \cosh(\beta H_B)
+ 2 \exp(\beta J/4) \cosh \Bigl(
             \beta \sqrt{J_1^2 + (J \Delta)^2}/2
                          \Bigr) \nonumber \\
V_4 \! \! \! &=& \! \! \! \exp(- \beta J/4) \sinh(\beta J_1/2 + \beta H_B)/2 \nonumber \\
V_5 \! \! \! &=& \! \! \! \exp(- \beta J/4) \sinh(\beta J_1/2 - \beta H_B)/2 \nonumber \\
V_6 \! \! \! &=& \! \! \! \exp(- \beta J/4) \sinh(\beta H_B)/2, \nonumber \\
V_7 \! \! \! &=& \! \! \!
\frac{2 J_1 \exp(\beta J/4)}{\sqrt{J_1^2 + (J \Delta)^2}}
\sinh \Bigl(\beta \sqrt{J_1^2 + (J \Delta)^2}/2 \Bigr). \nonumber
\label{a1}
\end{eqnarray}
b) Spin-1 chain:
\begin{eqnarray}
V_1 \! \! \! &=& \! \! \! 2 \exp(- \beta J) \cosh(\beta J_1 + 2 \beta H_B)
       + \exp(2 \beta J /3) W_1 + \nonumber \\
&& + 4 \cosh(\beta J_1/2 + \beta H_B) \cosh(\beta J \Delta) \nonumber \\
V_2 \! \! \! &=& \! \! \! 2 \exp(- \beta J) \cosh(\beta J_1 - 2 \beta H_B)
+ \exp(2 \beta J /3) W_1 + \nonumber \\
&& + 4 \cosh(\beta J_1/2 - \beta H_B) \cosh(\beta J \Delta) \nonumber \\
V_3 \! \! \! &=& \! \! \! 2 \exp(- \beta J) \cosh(2 \beta H_B)
+ \exp(2 \beta J /3) W_2 + \nonumber \\
&& + 4 \cosh(\beta H_B) \cosh
\Bigl( \beta \sqrt{J_1^2 + (2 J \Delta)^2}/2 \Bigr)       \nonumber \\
V_4 \! \! \! &=& \! \! \! \exp(- \beta J) \sinh(\beta J_1 + 2 \beta H_B)
 + \sinh(\beta J_1/2 + \beta H_B) \cosh(\beta J \Delta) \nonumber \\
V_5 \! \! \! &=& \! \! \! \exp(- \beta J) \sinh(\beta J_1 - 2 \beta H_B)
 + \sinh(\beta J_1/2 - \beta H_B) \cosh(\beta J \Delta) \nonumber \\
V_6 \! \! \! &=& \! \! \! \exp(- \beta J) \sinh(2 \beta H_B)
       + \sinh(\beta H_B) \cosh
\Bigl(\beta \sqrt{J_1^2 + (2 J \Delta)^2}/2 \Bigr)      \nonumber
\label{a2}
\end{eqnarray}
where the expressions $W_1$ and $W_2$ are given by:
\begin{eqnarray}
W_1 \! \! \! &=& \! \! \! \sum_{n=0}^2 \exp \Bigl \{- 2 \beta P_1 \cos[(\phi_1 + 2 \pi n)/3]
\Bigr \},
\nonumber \\
W_2 \! \! \! &=& \! \! \! \sum_{n=0}^2 \exp \Bigl \{- 2 \beta P_2 \cos
\bigl[ (\phi_2 + 2 \pi n)/3 \bigr]
\Bigr \},
\nonumber \\
P_1^2 \! \! \! &=& \! \! \! (J/3)^2 + 2 (J \Delta)^2/3,  \hspace{1.5cm}
P_2^2 = (J/3)^2 + 2 (J \Delta)^2/3 + J_1^2/3,    \nonumber \\
Q_1 \! \! \! &=& \! \! \! (J/3)^3 + J (J \Delta)^2/3,  \hspace{1.4cm}
Q_2 = (J/3)^3 + J (J \Delta)^2/3 - J_1^3,     \nonumber \\
\phi_1 \! \! \! &=& \! \! \! \arctan \Big( \sqrt{P_1^6 - Q_1^2}/Q_1 \Bigr),
\hspace{0.7cm}
\phi_2 = \arctan \Bigl( \sqrt{P_2^6 - Q_2^2}/Q_2 \Bigr).  \nonumber
\label{a3}
\end{eqnarray}

\newpage

{\bf Figure captions}
\begin{itemize}
\item [Fig.1]
Part of the doubly decorated mixed-spin chain.
The black circles denote the spin-$1/2$ Ising atoms of sublattice $A$
and the gray ones represent the decorating spin-$S$ Heisenberg atoms
of sublattice $B$. The ellipse demarcates a typical bond described by
the Hamiltonian $\hat {\cal H}_k$ introduced in Eq. (\ref{r2}).
\item [Fig.2]
Ground-state phase boundaries in the $\Delta$ - $H/J$ plane
for $J_1/J = 1.0$ and $2.0$.
\item [Fig.3]
Low-temperature ($k_B T / J = 0.01$) magnetization curves for:
a) $J_1/J = 1.0$ and $\Delta = 1.0$; b)-d) $J_1/J = 2.0$ and $\Delta =
0.5, 1.0, 1.5$. The solid and dotted (dashed) lines represent
the total magnetization per one site and the single-site
magnetization of the Ising (Heisenberg) sublattice, respectively.
\item [Fig.4]
Low-temperature behaviour ($k_B T / J = 0.01$) of several
correlation functions: $q_{hh}^{zz}$ and $q_{ii}^{zz}$ (dashed lines),
$q_{hh}^{xx}$ (dotted lines) and $q_{ih}^{zz}$ (solid lines).\\
a) plot of the correlation functions against the anisotropy
$\Delta$ for the system without external field; \\
b)-d) field-dependence of the correlation functions
depicted for $J_1/J = 2.0$ and various anisotropies
$\Delta = 0.5$, $1.0$ and $1.5$.
\item [Fig.5]
Thermal variations of the total magnetization for some typical
values of the external field, $J_1/J = 2.0$ and two selected
anisotropies $\Delta = 0.5$ and $1.5$, respectively.
\item [Fig.6]
Thermal behaviour of the system for $J_1/J = 2.0$,
$\Delta = 1.5$ and several values of external field. \\
a) entropy plot versus temperature; \\
b)-d) variations of the specific heat with temperature.
\item [Fig.7]
Ground-state phase diagram together with some typical examples
of low-temperature magnetization curves when the ratio $J_1/J = 1.0$; \\
a) ground-state phase diagram in the $\Delta-H/J$ plane; \\
b)-d) some typical examples of the low-temperature ($k_B T/J=0.001$)
magnetization curves for various anisotropies $\Delta$.
\item [Fig.8]
Low-temperature ($k_B T/J=0.001$) behaviour of several
correlation functions $q_{hh}^{zz}$ (dashed), $q_{hh}^{xx}$ (dotted),
$q_{ih}^{zz}$ (solid), $q_{ii}^{zz}$ (dashed-dotted),
quadrupolar momentum $\eta$ (dashed-dotted-dotted)
and staggered sublattice magnetization $m_B^s$ (solid line). \\
a) zero-field variations of relevant quantities with anisotropy $\Delta$; \\
b)-d) field-dependencies of correlations for selected values of $\Delta$.
\item [Fig.9]
Ground-state phase diagrams together with some typical examples
of low-temperature magnetization curves: \\
a)-b) ground-state phase diagram in the $\Delta-H/J$ plane for
$J_1/J = 0.5$ and $2.0$; \\
c)-d) some typical examples of the low-temperature ($k_B T/J=0.001$)
magnetization curves when $J_1/J = 2.0$ and $\Delta = 1.9$ and $2.05$,
respectively.
\item [Fig.10]
Thermal variations of the total magnetization for
$J_1/J = 1.0$ and two selected values of $\Delta$.
\end{itemize}

\end{document}